\documentclass[
12pt,
preprint,preprintnumbers,nofootinbib,
groupedaddress,superscriptaddress,amsmath,amssymb]{revtex4}
\usepackage{graphicx}
\usepackage{dcolumn}
\usepackage{bm}
\usepackage{amssymb}
\usepackage{amsmath}
\usepackage{epsfig}    
\usepackage{color}
\usepackage{hhline}

\def\be{\begin{equation}}
\def\ee{\end{equation}}
\newcommand{\bea}{\begin{eqnarray}}
\newcommand{\eea}{\end{eqnarray}}
\newcommand{\nn}{\nonumber}

\numberwithin{equation}{section}

\begin{document}
{\begin{flushright}{KIAS-P19048, APCTP Pre2019 - 022}\end{flushright}}

\title{A modular $A_4$ symmetric scotogenic model}

\author{Takaaki Nomura}
\email{nomura@kias.re.kr}
\affiliation{School of Physics, KIAS, Seoul 02455, Republic of Korea}

\author{Hiroshi Okada}
\email{hiroshi.okada@apctp.org}
\affiliation{Asia Pacific Center for Theoretical Physics (APCTP) - Headquarters San 31, Hyoja-dong,
Nam-gu, Pohang 790-784, Korea}
\affiliation{Department of Physics, Pohang University of Science and Technology, Pohang 37673, Republic of Korea}

\author{Oleg Popov}
\email{opopo001@ucr.edu}
\affiliation{Institute of Convergence Fundamental Studies, \\ Seoul National University of Science and Technology, \\Seoul 139-743, Korea}

\date{\today}

\begin{abstract}
{
We propose a minimal extention of the Standard Model where neutrino masses are generated radiatively at one-loop level via Scotogenic scanario. The model is augmented with $A_4$ modular symmetry as a scotogenic and flavor symmetry. With minimal number of parameters, the model makes predictions for neutrino oscillation data, Majorana and Dirac phases, dark matter characteristics, and neutrinoless double beta decay.
}
\end{abstract}
\maketitle
\newpage

\section{Introduction}

Particle physics experiments and observations have been successfully confirming the standard model (SM) of particle physics.
On the other hand, there are some issues indicating an existence of physics beyond the SM such as existence of dark matter (DM), non-zero tiny neutrino masses and origin of flavor structure.
In describing these issues, symmetry would play an important role like guaranteeing stability of DM, forbidding neutrino mass at tree level and restricting flavor structure.
It is thus interesting to construct a model of physics beyond the SM adopting a new symmetry.

Modular flavor symmetries have been recently proposed by~\cite{Feruglio:2017spp, deAdelhartToorop:2011re}
to provide more predictions to the quark and lepton sector due to Yukawa couplings with a representation of a group.  
Their typical groups are found in basis of  the $A_4$ modular group \cite{Feruglio:2017spp, Criado:2018thu, Kobayashi:2018scp, Okada:2018yrn, Nomura:2019jxj, Okada:2019uoy, deAnda:2018ecu, Novichkov:2018yse, Nomura:2019yft, Okada:2019mjf}, $S_3$ \cite{Kobayashi:2018vbk, Kobayashi:2018wkl, Kobayashi:2019rzp, Okada:2019xqk}, $S_4$ \cite{Penedo:2018nmg, Novichkov:2018ovf, Kobayashi:2019mna}, $A_5$ \cite{Novichkov:2018nkm, Ding:2019xna}, larger groups~\cite{Baur:2019kwi}, multiple modular symmetries~\cite{deMedeirosVarzielas:2019cyj}, and double covering of $A_4$~\cite{Liu:2019khw} in which  masses, mixings, and CP phases for quark and lepton are predicted.~\footnote{Several reviews are helpful to understand the whole idea~\cite{Altarelli:2010gt, Ishimori:2010au, Ishimori:2012zz, Hernandez:2012ra, King:2013eh, King:2014nza, King:2017guk, Petcov:2017ggy}.}
Also, a systematic approach to understand the origin of CP transformations has been recently achieved by ref.~\cite{Baur:2019iai}. 
{In particular a model with modular $A_4$ symmetry is discussed in ref.~\cite{Nomura:2019jxj,Okada:2019mjf} where neutrino mass is generated at one-loop level.
For model in ref.~\cite{Nomura:2019jxj}, inert doublet and singlet scalar fields are introduced to generate neutrino mass using modular form of weight 2 which is a triplet of $A_4$ and the basis for constructing other modular forms with higher weight. On the other hand, for model in ref.~\cite{Okada:2019mjf}, vector-like leptons are introduced to get positive contribution to muon $g-2$ and triplet modular form with weight 4 is used. 

In this paper, we apply modular $A_4$ symmetry in minimal Scotogenic model~\cite{Ma:2006km} in which neutrino mass is generated at one-loop level and DM candidates are contained. 
We find that the minimal Scotogenic model can be realized by use of modular forms with higher weight which are constructed by that of weight 2. 
Thus field contents and the structure of model are much simpler than previous models~\cite{Nomura:2019jxj,Okada:2019mjf}.
In our construction the right-handed neutrinos are introduced as a triplet of $A_4$ assigning modular weight $-k=-1$.
Also, non-zero modular weight is assigned to inert Higgs doublet as $-k=-3$. }
Interestingly we find that additional $Z_2$ symmetry is not necessary to realize structure of Scotogenic model due to the nature of modular form.
Then numerical analysis for neutrino mass matrix is carried out to show predictions of our model as a result of modular $A_4$ symmetry.

{Manuscript} is organized as follows.
{In Sec.~\ref{sec:realization}, we give our model set up under $A_4$ modular symmetry.
We discuss right-handed neutrino mass spectrum, lepton flavor violation (LFV) and generation of the active neutrino mass at one-loop level in Sec.~\ref{sec:analysis}. Numerial analysis is presented in Sec.~\ref{sec:num_analysis}. Finally, we conclude and discuss in Sec.~\ref{sec:conclusion}.}

\begin{center} 
\begin{table}[tb]
\begin{tabular}{||c||c|c|c||c|c||}\hline
&\multicolumn{3}{c||}{Fermions} & \multicolumn{2}{c||}{Bosons} \\ \hline \hline
  & ~$(\bar L_{L_e},\bar L_{L_\mu},\bar L_{L_\tau})$~ & ~$(e_{R_e},e_{R_{\mu}},e_{R_{\tau}})$ ~ & ~$N_{R}$~ & ~$H$~  & ~$\eta^*$~
  \\\hline 
 $SU(2)_L$ & $\bm{2}$     & $\bm{1}$  & $\bm{1}$ & $\bm{2}$ & $\bm{2}$    \\\hline 
$U(1)_Y$ & $\frac12$  & $-1$ & $0$  & $\frac12$  & -$\frac12$      \\\hline
 $A_4$ & ${1,1',1''}$ & ${1,1'',1'}$ & $3$ & $1$ & $1$ \\\hline
 $-k$ & $0$ & $0$ & $-1$ & $0$ & $-3$   \\\hline
\end{tabular}
\caption{Fermionic and bosonic field content of the model and their charge assignments under $SU(2)_L\times U(1)_Y\times A_4$ in the lepton and boson sector, where $-k$ is the number of modular weight 
and the quark sector is the same as the SM.}
\label{tab:fields}
\end{table}
\end{center}

\begin{center} 
\begin{table}[tb]
\begin{tabular}{|c||c|c|c|c|c|c|}\hline
 &\multicolumn{3}{c|}{Couplings}  \\\hline\hline
  & ~$Y^{(6)}_{\bf1}$~& ~$Y^{(2)}_{\bf3}$~ & ~$Y^{(4)}_{\bf3}$~    \\\hline 
{ $A_4$} & ${\bf1}$ & ${\bf3}$& ${\bf3}$      \\\hline
 $-k$ & $6$ & $2$ & $4$     \\\hline
\end{tabular}
\caption{Modular weight assignments for Yukawas.}
\label{tab:couplings}
\end{table}
\end{center}

\section{Model}
\label{sec:realization}
{In this section we introduce our model, which is based on $A_4$ modular symmetry. }
{Leptonic and scalar fields of the model and their representations under $A_4$ symmetry} and modular weights are given by Tab.~\ref{tab:fields}, while the ones of Yukawa couplings are given by Tab.~\ref{tab:couplings}.
Under these symmetries, we write renormalizable Lagrangian as follows:
\begin{align}
-{\cal L}_{Lepton} &=
\sum_{\ell=e,\mu,\tau}y_\ell \bar L_{L_\ell} H e_{R_\ell}\nn\\
&+\alpha_\nu \bar L_{L_e} (Y^{(4)}_{\bf 3}\otimes N_{R})_{\bf1}\tilde\eta
+\beta_\nu \bar L_{L_\mu}(Y^{(4)}_{\bf 3}\otimes N_{R})_{\bf1''}\tilde\eta
+\gamma_\nu \bar L_{L_\tau}(Y^{(4)}_{\bf 3} \otimes N_{R})_{\bf1'}\tilde\eta\nn\\
&+ {M_1} (Y^{(2)}_{\bf 3} \otimes\bar N^C_{R}\otimes N_{R})
+ {\rm h.c.}, 
\label{eq:lag-lep}
\end{align}
where $\tilde\eta\equiv i\sigma_2 \eta^*$, $\sigma_2$ being second Pauli matrix, and charged-lepton matrix is diagonal thanks to the unique representation of $A_4$.

{The full symmetry of the leptonic sector of the model is $SU(2)_L \times U(1)_Y \times A_4$, where the $\mathbb{Z}_2$ symmetry of~\cite{Ma:2006km} is replaced with modular symmetry $\Gamma_3\simeq A_4$. $A_4$ serves three purposes: flavor symmetry, scotogenic symmetry, and dark matter stabilizing symmetry.}

The basis of modular form is the one with weight 2, $Y^{(2)}_{\bf 3} =(y_{1},y_{2},y_{3})$, transforming as a triplet of $A_4$ whose components  are written in terms of Dedekind eta-function  $\eta(\tau)$ and its derivative $\eta'(\tau)$~\cite{Feruglio:2017spp}:
\begin{eqnarray} 
\label{eq:Y-A4}
y_{1}(\tau) &=& \frac{i}{2\pi}\left( \frac{\eta'(\tau/3)}{\eta(\tau/3)}  +\frac{\eta'((\tau +1)/3)}{\eta((\tau+1)/3)}  
+\frac{\eta'((\tau +2)/3)}{\eta((\tau+2)/3)} - \frac{27\eta'(3\tau)}{\eta(3\tau)}  \right), \nonumber \\
y_{2}(\tau) &=& \frac{-i}{\pi}\left( \frac{\eta'(\tau/3)}{\eta(\tau/3)}  +\omega^2\frac{\eta'((\tau +1)/3)}{\eta((\tau+1)/3)}  
+\omega \frac{\eta'((\tau +2)/3)}{\eta((\tau+2)/3)}  \right) , \label{eq:Yi} \\ 
y_{3}(\tau) &=& \frac{-i}{\pi}\left( \frac{\eta'(\tau/3)}{\eta(\tau/3)}  +\omega\frac{\eta'((\tau +1)/3)}{\eta((\tau+1)/3)}  
+\omega^2 \frac{\eta'((\tau +2)/3)}{\eta((\tau+2)/3)}  \right)\,.
\nonumber
\end{eqnarray}
%
Here the overall coefficient in Eq. (\ref{eq:Yi}) is one possible choice; it cannot be uniquely determined. 
Then, any modular forms of higher weight are constructed by the products of $Y^{(2)}_{\bf 3}$ using multiplication rules of $A_4$ representations, and one finds the following higher weight modular forms:
\begin{align}
&Y^{(6)}_{\bf1}=y^3_1 + y^3_2+y^3_3-3y_1y_2y_3,\quad
Y^{(4)}_{\bf3}=
\left[\begin{array}{c}
y^2_1-y_2y_3 \\ 
y^2_3-y_1y_2 \\ 
y^2_2-y_1y_3 \\ 
\end{array}\right].
\end{align}

Higgs potential of our model is equivalent to the potential of the Scotogenic model~\cite{Ma:2006km} without loss of generality, where a quartic coupling that plays {the} role in generating the nonzero neutrino masses is given by $Y^{(6)}_{\bf1}(H^\dag \eta)^2$ term.  The $(H^\dag \eta)$ term that was forbidden in the inert two Higgs doublet model (2HDM) by $\mathbb{Z}_2$ invariance is now forbidden by modular invariance via $A_4$. 
{This is due to the fact that modular form with odd number of modular weight does not exist; here the oddness of modular weight play a role of odd parity under $Z_2$ symmetry as shown in ref.~\cite{Nomura:2019jxj}. }

The right-handed neutrino mass matrix is given by
\begin{align}
M_N&= \frac{M_1}3
\left[\begin{array}{ccc}
2y_{1} & -y_{3} & -y_{2} \\ 
-y_{3} & 2y_{2} & -y_{1} \\ 
-y_{2} & -y_{3,1} &2 y_{3} \\ 
\end{array}\right] .
\end{align}
Then, the Majorana mass matrix is diagonalized by an unitary matrix as $D_N\equiv U M_N U^T$, and their mass eigenstates are defined by $\psi_R$, where $N_R=U^T\psi_R$. 
  
The Dirac Yukawa matrix is given by
\begin{align}
y_\eta &=
\left[\begin{array}{ccc}
\alpha_\nu &0 & 0 \\ 
0 & \beta_\nu  &0   \\ 
0& 0 & \gamma_\nu  \\ 
\end{array}\right]
\left[\begin{array}{ccc}
y'_{1} & y'_{3} & y'_{2} \\ 
y'_{3} & y'_{2} & y'_{1} \\ 
y'_{2} & y'_{1} & y'_{3} \\ 
\end{array}\right],\label{eq:mn}
\end{align}
where $Y^{(4)}_{\bf3}\equiv (y'_{1},y'_{2},y'_{3})$,
and we impose the perturbative limit {${\rm Max}[y_{\eta}]\lesssim\sqrt{4\pi}$} in the numerical analysis.
\section{Analysis}
\label{sec:analysis}

{In this section we analyze lepton flavor violation and neutrino mass formulating analytic forms of branching ratio (BR) of $\ell_i \to \ell_j \gamma$ process and neutrino mass matrix.}

{\it  Charged lepton flavor violating (cLFV) processes} arise from Yukawa interactions associated with $y_\eta$ coupling as in~\cite{Baek:2016kud}.
Considering the mixing matrix of $N_R$, we obtain the BRs such that
\begin{align}
&{\rm BR}(\ell_i\to\ell_j\gamma)\approx\frac{48\pi^3\alpha_{em}C_{ij}}{G_F^2 (4\pi)^4}
\left|\sum_{\alpha=1-3}Y_{\eta_{j\alpha}} Y^\dag_{\eta_{\alpha i}} F(D_{N_\alpha},m_{\eta^\pm})\right|^2,\\
&F(m_a,m_b)\approx\frac{2 m^6_a+3m^4_am^2_b-6m^2_am^4_b+m^6_b+12m^4_am^2_b\ln\left(\frac{m_b}{m_a}\right)}{12(m^2_a-m^2_b)^4},
\end{align}
where $Y_\eta\equiv y_\eta U^T$, $C_{21}=1$, $C_{31}=0.1784$, $C_{32}=0.1736$, $\alpha_{em}(m_Z)=1/128.9$, and $G_F=1.166\times10^{-5}$ GeV$^{-2}$.
The experimental upper bounds are given by~\cite{TheMEG:2016wtm, Aubert:2009ag,Renga:2018fpd}
\begin{align}
{\rm BR}(\mu\to e\gamma)\lesssim 4.2\times10^{-13},\quad 
{\rm BR}(\tau\to e\gamma)\lesssim 3.3\times10^{-8},\quad
{\rm BR}(\tau\to\mu\gamma)\lesssim 4.4\times10^{-8},
\label{eq:lfvs-cond}
\end{align}
which will be imposed in our numerical calculation as constraints.
 %

{\it Neutrino mass matrix}  at one-loop level can be derived as
\begin{align}
&m_{\nu_{ij}}\approx \sum_{\alpha=1-3}\frac{Y_{\eta_{i\alpha}} {D_{N}}_\alpha Y^T_{\eta_{\alpha j}}}{(4\pi)^2}
\left(\frac{m_R^2}{m_R^2-{ D^2_N}_{\alpha}}\ln\left[\frac{m_R^2}{{ D^2_N}_{\alpha}}\right]
-
\frac{m_I^2}{m_I^2-{D^2_N}_{\alpha}}\ln\left[\frac{m_I^2}{{ D^2_N}_{\alpha}}\right]
\right),
\end{align}
where $m_I$ and $m_R$ are, respectively, {masses of imaginary and real parts of} neutral component in $\eta$.
Then, the neutrino mass matrix is diagonalized by the PMNS unitary matrix, $U_{PMNS}$, as $U_{PMNS}m_\nu U^T_{PMNS}=$diag($m_{\nu_1},m_{\nu_2},m_{\nu_3}$)$\equiv D_\nu$, since the charged-lepton mass matrix is diagonal in our model.
Note that the constraint for sum of neutrino mass Tr$[D_{\nu}]\lesssim 0.12$ eV is given by the data of recent cosmological observations~\cite{Aghanim:2018eyx}.
Each of mixing angle is given in terms of the component of $U_{PMNS}$ as follows:
\begin{align}
\sin^2\theta_{13}=|(U_{PMNS})_{13}|^2,\quad 
\sin^2\theta_{23}=\frac{|(U_{PMNS})_{23}|^2}{1-|(U_{PMNS})_{13}|^2},\quad 
\sin^2\theta_{12}=\frac{|(U_{PMNS})_{12}|^2}{1-|(U_{PMNS})_{13}|^2}.
\end{align}
In addition, the effective mass for the neutrinoless double beta decay is given by
\begin{align}
m_{ee}=|D_{\nu_1} \cos^2\theta_{12} \cos^2\theta_{13}+D_{\nu_2} \sin^2\theta_{12} \cos^2\theta_{13}e^{i\alpha_{21}}
+D_{\nu_3} \sin^2\theta_{13}e^{i(\alpha_{31}-2\delta_{CP})}|,
\end{align}
where its observed value could be tested by KamLAND-Zen in future~\cite{KamLAND-Zen:2016pfg}.

To carry out numerical analysis, we derive several relations between the normalized neutrino mass matrix and our parameters as follows:
\begin{align}
&\tilde m_{\nu_{ij}}\equiv \frac{m_{\nu_{ij}}}{k_3}\approx 
\frac{1}{(4\pi)^2}\sum_{\alpha=1-3}
Y_{\eta_{i\alpha}} \tilde k_\alpha Y^T_{\eta_{\alpha j}},\quad \tilde k_\alpha \equiv \frac{k_\alpha}{k_3} ,\nn\\
 k_\alpha&\equiv {D_N}_{\alpha}
\left(\frac{m_R^2}{m_R^2-{ D^2_N}_{\alpha}}\ln\left[\frac{m_R^2}{{D^2_N}_{\alpha}}\right]
-
\frac{m_I^2}{m_I^2-{D^2_N}_{\alpha}}\ln\left[\frac{m_I^2}{{ D^2_N}_{\alpha}}\right]\right)\nn\\
&\approx 
 {D_N}_{\alpha} \Delta m^2 \left(\frac{ { {D^2_N}_{\alpha} - m_R^2 + {D^2_N}_{\alpha} \ln\left(\frac{m_R^2}{ {D^2_N}_{\alpha}}\right)}}{({ D^2_N}_{\alpha} - m_R^2)^2}\right),
\label{eq:norm-nu}
\end{align}
where the last line is the first order approximation in terms of the small mass difference between $m_R^2$ and $m_I^2$ defined by
$m_R^2-m_I^2=\Delta m^2$.~\footnote{Advantage of this approximation is that $\tilde k_\alpha$ does not depend on $\Delta m$.}
Then, the normalized neutrino mass eigenvalues are written in terms of neutrino mass eigenvalues; diag$(\tilde m_{\nu_1}^2,\tilde m_{\nu_2}^2,\tilde m_{\nu_3}^2)={\rm diag}(m_{\nu_1}^2,m_{\nu_2}^2,m_{\nu_3}^2)/k_3^2$. 
 It is then found that $k_3^2$ is given by
\begin{align}
k_3^2=\frac{\Delta m^2_{\rm atm}}{\tilde m_{\nu_3}^2 - \tilde m_{\nu_1}^2},
\label{eq:k}
\end{align}
where normal hierarchy is assumed here and $\Delta m^2_{\rm atm}$ is the atmospheric neutrino mass difference square.
Thus, comparing Eq.(\ref{eq:norm-nu}) and Eq.(\ref{eq:k}), we can rewrite $\Delta m^2$ by other parameters as follows:
\begin{align}
\Delta m^2 \approx
 k_3 \left(\frac{ {D_N}_{3} \left[{ {D^2_N}_{3} - m_R^2 + {D^2_N}_{3} \ln\left(\frac{m_R^2}{ {D^2_N}_{3}}\right)}\right] }
{({ D^2_N}_{3} - m_R^2)^2}\right)^{-1} .
 \label{eq:dm}
\end{align}
 The solar neutrino mass difference square is also found as 
 \begin{align}
\Delta m^2_{\rm sol}=\Delta m^2_{\rm atm} \frac{\tilde m_{\nu_2}^2 - \tilde m_{\nu_1}^2}{\tilde m_{\nu_3}^2 - \tilde m_{\nu_1}^2},
\label{eq:m2sol}
\end{align}
In numerical analysis, we require this value to be within the experimental result, while we take $\Delta m^2_{\rm atm}$ as an input parameter.

%
\section{Numerical analysis}
\label{sec:num_analysis}
We show numerical analysis to satisfy all of the constraints that we discussed above,
where we assume $m_R\approx m_I\approx m_{\eta^\pm}$ to avoid the constraint of oblique parameters.
Also, we impose the recent cosmological data; Tr$[D_{\nu}]\lesssim 0.12$ eV.~\footnote{If this constraint is removed, another allowed range can be found.}
Then, we provide the experimentally allowed ranges for neutrino mixings and mass difference squares at 3$\sigma$ range~\cite{Esteban:2018azc} as follows:
\begin{align}
&\Delta m^2_{\rm atm}=[2.431-2.622]\times 10^{-3}\ {\rm eV}^2,\
\Delta m^2_{\rm sol}=[6.79-8.01]\times 10^{-5}\ {\rm eV}^2,\\
&\sin^2\theta_{13}=[0.02044-0.02437],\ 
\sin^2\theta_{23}=[0.428-0.624],\ 
\sin^2\theta_{12}=[0.275-0.350].\nn
\end{align}
{The range of absolute values in three dimensionless parameters $\alpha_\nu,\beta_\nu, \gamma_\nu$ are taken to be $[0.1-1]$,}
while the mass parameters $M_1$ is of the order $100$ TeV.
{We also choose $m_R = 534 \pm 8.5$ GeV and $m_I = \sqrt{m_R^2 + \Delta m^2 }$ for inert scalar mass.}

\begin{figure}[tb]\begin{center}
\includegraphics[width=80mm]{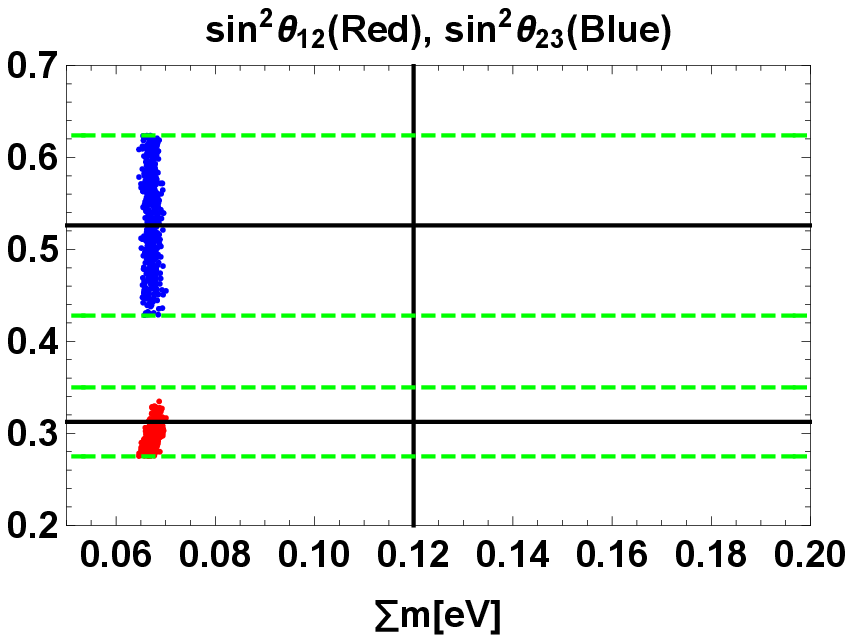}
\includegraphics[width=80mm]{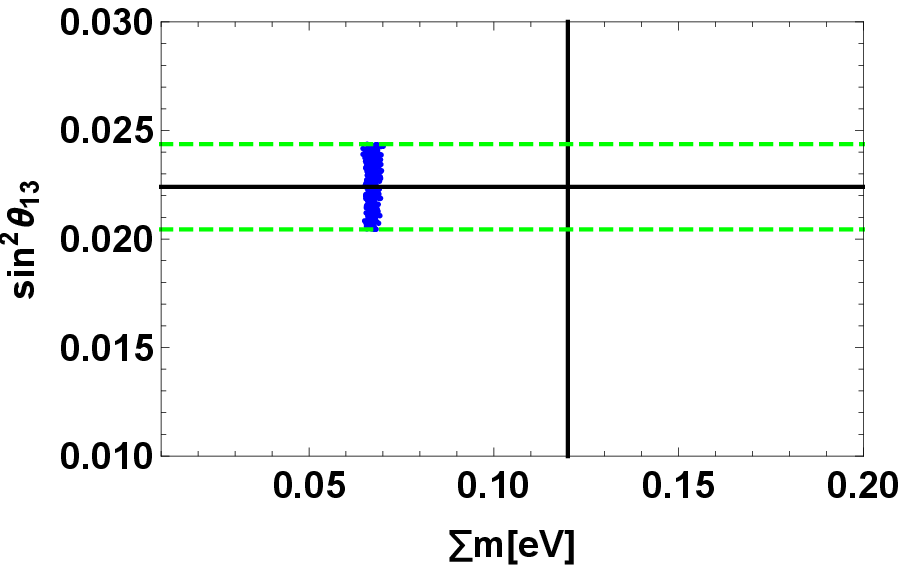}
\caption{The sum of neutrino masses $\sum m(\equiv \sum D_\nu)$ versus $\sin^2\theta_{12}$(red color) and $\sin^2\theta_{23}$(blue color) in the left figure and  $\sin^2\theta_{13}$(blue color) in the right figure.
Here, the horizontal black solid lines are the best fit values, the green dotted lines show 3$\sigma$ range,
and the vertical black line shows upper bound on the cosmological data as shown in the neutrino section.}   
\label{fig:1}\end{center}\end{figure}

Fig.~\ref{fig:1} shows the sum of neutrino masses $\sum m(\equiv \sum D_\nu)$ versus $\sin^2\theta_{12}$(red color) and
$\sin^2\theta_{23}$(blue color) in the left figure and  $\sin^2\theta_{13}$(blue color) in the right figure.
Here, the horizontal black solid lines are the best fit values, the green dotted lines show 3$\sigma$ range,
and the vertical black line shows upper bound on the cosmological data as shown in the neutrino section.
It suggests that all the three mixings run over the experimental ranges, while 
$\sum m$ is allowed by the narrow range of [0.065-0.070] eV that is always below the cosmological bound 0.12 eV.

\begin{figure}[tb]\begin{center}
\includegraphics[width=80mm]{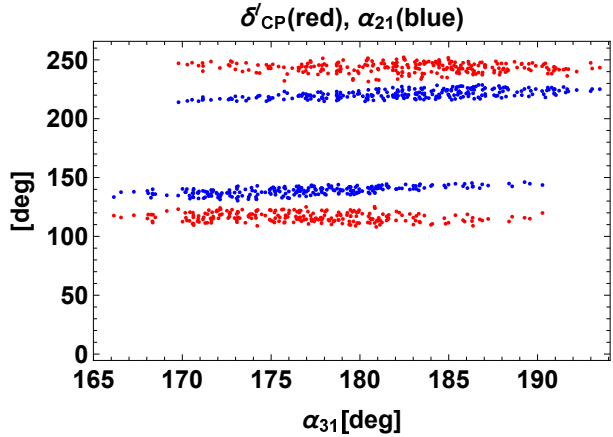}
\caption{Phases of $\delta^\ell_{CP}$(red color) and  $\alpha_{21}$(blue color) in terms of $\alpha_{31}$.}   
\label{fig:2}\end{center}\end{figure}

Fig.~\ref{fig:2} shows phases of $\delta^\ell_{CP}$(red color) and  $\alpha_{21}$(blue color) in terms of $\alpha_{31}$.
{This figure implies that Dirac CP is allowed by the range [100-120, 230-250] [deg], $\alpha_{21}$ is [130-150, 210-230] [deg], and
 $\alpha_{31}$ is [165-190] [deg].}

\begin{figure}[tb]\begin{center}
\includegraphics[width=80mm]{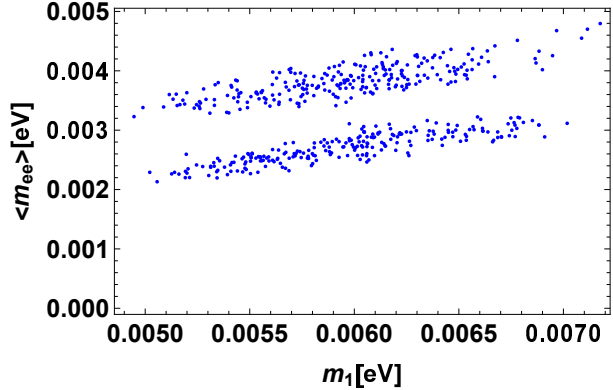}
\caption{The lightest neutrino mass versus {the effective mass for the} neutrinoless double beta decay.}   
\label{fig:3}\end{center}\end{figure}

\begin{figure}[tb]\begin{center}
\includegraphics[width=80mm]{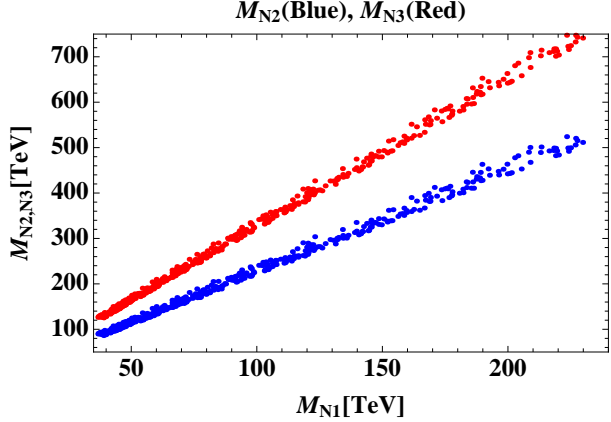}
\caption{The correlations among masses of $N_i$ where red(blue) color points correspond to $M_{N_1}-M_{N_{3(2)}}$ correlation.}   
\label{fig:4}\end{center}\end{figure}

\begin{figure}[tb]\begin{center}
\includegraphics[width=50mm]{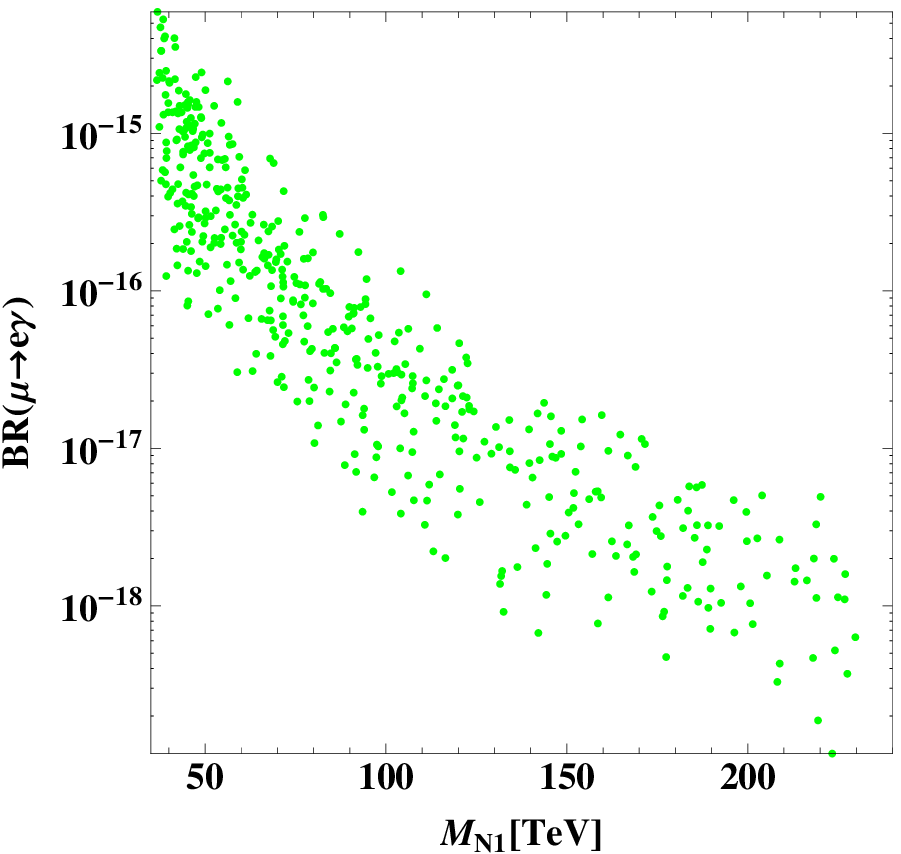}
\includegraphics[width=50mm]{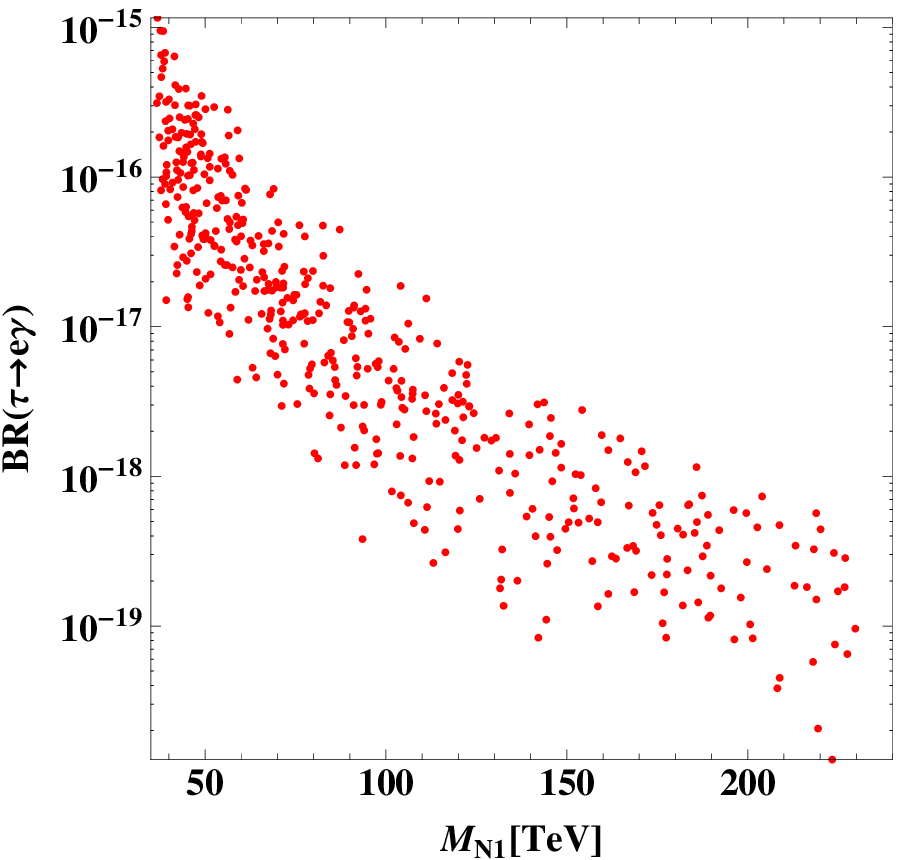}
\includegraphics[width=50mm]{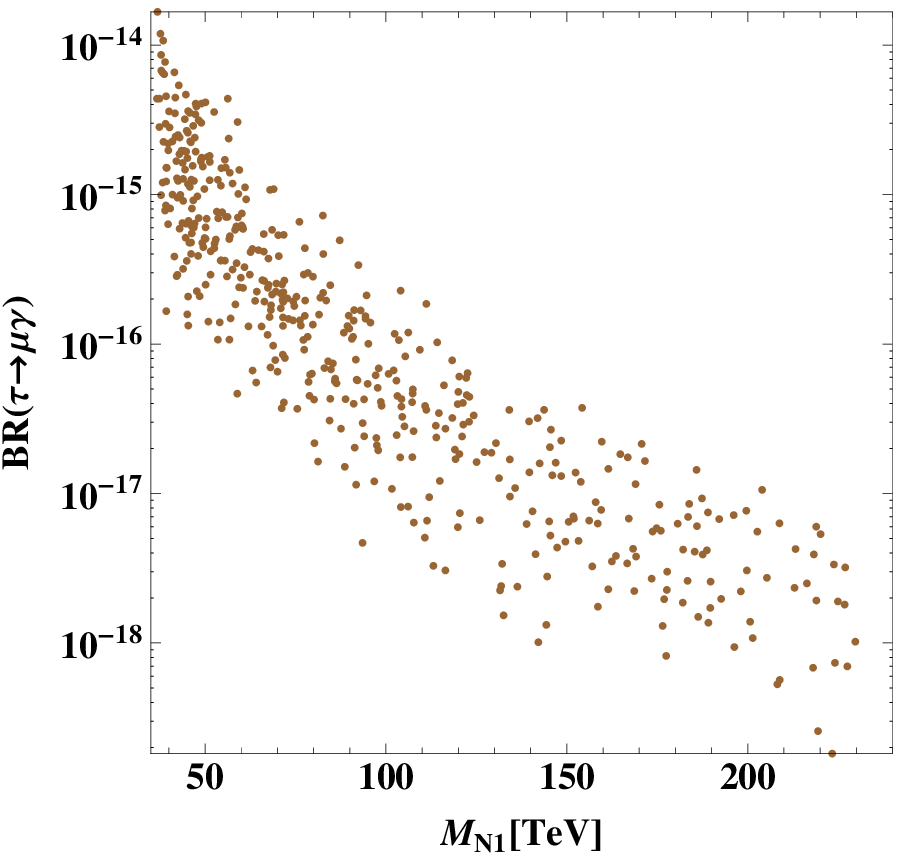}
\caption{The BRs for cLFV processes $\ell \to \ell' \gamma$ as functions of the mass of $N_1$.}   
\label{fig:5}\end{center}\end{figure}

Fig.~\ref{fig:3} demonstrates the lightest neutrino mass versus {the effective mass for the} neutrinoless double beta decay.
It suggests that $0.0049\lesssim m_1\lesssim0.0072$ eV and $0.002\lesssim\langle m_{ee}\rangle\lesssim0.005$ eV.
Another remarks are in order:
 \begin{enumerate}
\item
The typical region of modulus $\tau$ is found in rather narrow modular field space as 0.43\ $\lesssim\ $Re$[\tau]\lesssim$\ 0.45 and 0.65\ $\lesssim\ $Im$[\tau]\lesssim$\ 0.67.
\item The  Majorana mass eigenvalues are in the range of 
\[D_{N_1}=[40-230]\ {\rm TeV},\quad D_{N_2}=[80-520]\  {\rm TeV},\quad D_{N_3}=[100-750] \ {\rm TeV}.\]
{We also show correlation among the mass eigenvalues in Fig.~\ref{fig:4}. 
Note that the mass scale is larger than the previous model in ref.~\cite{Nomura:2019jxj} which is due to simpler loop structure requiring heavier masses of $N_i$.}
\item 
Typical scale of cLFVs tends to be very small in our analyses as shown in Fig.~\ref{fig:5}, therefore following upper bounds are realized:
\[{\rm BR}(\mu\to e\gamma)\lesssim6.0\times10^{-15}, \quad {\rm BR}(\tau\to e\gamma)\lesssim1.2\times10^{-15},\quad {\rm BR}(\tau\to \mu\gamma)\lesssim1.7\times10^{-14}.\]
{These values are smaller than the models in ref.~\cite{Nomura:2019jxj,Okada:2019mjf} which is due to heavier mass scale of $N_i$. }
 \end{enumerate}

{Note that DM candidate in our scenario is inert scalar $\eta_R$ since $N_1$ is much heavier.
The mass value of $m_R \sim 530$ GeV can accommodate with observed relic density of DM via gauge interaction taking into account coannihilation processes~\cite{Arhrib:2013ela}.
We also need to assume small Higgs portal coupling to avoid DM direct detection constraints. 
In principle, our DM phenomenology is the same as of the canonical inert Higgs doublet model and we do not discuss it in this paper.
}

\section{Conclusion and discussion}
\label{sec:conclusion}

We have studied a model based on modular $A_4$ symmetry in which neutrino masses are generated radiatively at one-loop level.
{The minimal Scotogenic scenario can be realized using modular form with higher weight where we have much simpler field contents
compared to previous modular $A_4$ radiative neutrino mass generation model with the modular form of lowest weight given in ref.~\cite{Nomura:2019jxj}. }
The modular $A_4$ symmetry plays a role of restricting interactions, generating neutrino mass and stabilizing DM candidate.
We have formulated lepton flavor violation and neutrino mass matrix in the model.
Then numerical analysis has been carried out to find prediction of our scenario.
In our numerical analyses, we have highlightend several remarks as follows: 
 \begin{enumerate}
\item
Three mixings cover all the experimental results by 3$\sigma$ interval, but the sum of neutrino masses are
the narrow range [0.065,0.0070] eV that is below the upper bound of cosmological data of 0.12 {eV}.
\item
{Dirac CP is allowed by the range [100-120, 230-250] [deg], $\alpha_{21}$ is [130-150, 210-230] [deg], and
 $\alpha_{31}$ is [165-190] [deg].  }
%
\item 
We found the following regions;
$0.0049\lesssim m_1\lesssim0.0072$ eV and $0.002\lesssim\langle m_{ee}\rangle\lesssim0.005$ eV {which can be seen from Fig.~\ref{fig:3}.}
 \end{enumerate}
These predictions will be tested in the near future. 
{In fact, comparing with previous one-loop models with modular $A_4$ in ref.~\cite{Nomura:2019jxj, Okada:2019mjf}, 
we find prediction for CP-phases are different although that for neutrino mass and mixing are similar.
Thus it would be possible to distinguish these models by measuring CP-phase such as Dirac phase $\delta^l_{CP}$ in future experiments.}
The DM candidate in our scenario is inert scalar boson and its mass is chosen to be $\sim 530$ GeV. 
DM phenomenology is the same as of the canonical inert Higgs doublet model and 
our DM mass value can accommodate the observed relic density of DM via gauge interaction taking into account coannihilation processes. 
We also require small Higgs portal coupling to avoid DM direct detection constraints, which can be easily realized by choosing parameters in the potential. \\

\section*{Acknowledgments}
\vspace{0.5cm}
{\it
This research was supported by an appointment to the JRG Program at the APCTP through the Science and Technology Promotion Fund and Lottery Fund of the Korean Government. This was also supported by the Korean Local Governments - Gyeongsangbuk-do Province and Pohang City (H.O.). 
H. O. is sincerely grateful for the KIAS member, and log cabin at POSTECH to provide nice space to come up with this project. OP is supported by the National Research Foundation of Korea Grants No. 2017K1A3A7A09016430 and No. 2017R1A2B4006338.}


\end{document}